\documentclass[preprint2]{aastex}
\pdfoutput=1

\usepackage{amsmath}
\usepackage{color}

\newcommand{\unit}[1]{~#1}
\newcommand{\Msol}{M$_{\odot}$}

\newcommand{\snaprow}[3]{
	\hspace*{0.8cm}\includegraphics[height=0.27\textwidth]{#1}
	\includegraphics[height=0.27\textwidth]{#2}
	\includegraphics[height=0.27\textwidth]{#3}\\
}
\newcommand{\evolrow}[1]{
	\includegraphics[height=0.294\textwidth]{#1}
}

\begin{document}

\shorttitle{Flybys and Bar Formation}

\title{Bar Formation from Galaxy Flybys}

\author{Meagan Lang\altaffilmark{1}, Kelly Holley-Bockelmann\altaffilmark{1,\,2}, \& Manodeep Sinha\altaffilmark{1}}

\altaffiltext{1}{Department of Physics and Astronomy, Vanderbilt
University, Nashville, TN email: {\tt meagan.lang@vanderbilt.edu}} 

\altaffiltext{2}{Fisk University, Department
of Physics, Nashville, TN email:{\tt k.holley@vanderbilt.edu}}

\begin{abstract}
Recently, both simulations and observations have revealed that flybys - fast, one-time interactions between two galaxy halos - are surprisingly common, nearing/comparable to galaxy mergers. Since these are rapid, transient events with the closest approach well outside the galaxy disk, it is unclear if flybys can transform the galaxy in a lasting way. We conduct collisionless N-body simulations of three co-planer flyby interactions between pure-disk galaxies to take a first look at the effects flybys have on disk structure, with particular focus on stellar bar formation. We find that some flybys are capable of inciting a bar with bars forming in both galaxies during our 1:1 interaction and in the secondary during our 10:1 interaction. The bars formed have ellipticities $\gtrsim0.5$, sizes on the order of the host disk's scale length, and persist to the end of our simulations, $\sim$5\unit{Gyr} after pericenter. The ability of flybys to incite bar formation implies that many processes associated with secular bar evolution may be more closely tied with interactions than previously though.
\end{abstract}

\keywords{galaxies: interactions --- galaxies: flybys --- galaxies: bars}

\section{Introduction}\label{S_intro}
While interacting galaxies bind together and eventually coalesce during a merger, more energetic encounters allow the two galaxies to disconnect and separate forever. These flybys generate a short, but intense, perturbation in both galaxies. Linear perturbation theory predicts that such an impulse should be similar in amplitude to that excited by a minor merger \citep{Vesperini2000} and therefore may transform galaxies in similar ways. 

For example, numerical simulations have shown that, if a disk is kinematically cool enough, even a small perturbation can induce a stellar bar \citep{Noguchi1987,Gerin1990,Miwa1998,Berentzen2004} and this is supported by bars triggered during minor merger simulations \citep{Noguchi1987,Steinmetz2002,Berentzen2004,Dubinski2008}. Almost at odds with this, simulations have shown that minor mergers can also destroy bars \citep{RomanoDiaz2008}. Therefore, the similar impulse induced by flybys could play a role in both the creation and destruction of bars.

Bars drive significant galaxy evolution in the form of angular momentum exchange \citep{Lynden-Bell1972,Tremaine1984,Weinberg1985,Athanassoula2002,Holley-Bockelmann2005}, gas inflow \citep{Hernquist1995}, nuclear star formation \citep{Younger2008}, and even super-massive black hole (SMBH) growth \citep{Hopkins2010b}. If flybys can excite or destroy bars, we may observe large-scale evolution beyond what is expected through hierarchical growth. In particular, bar-induced gas inflow has been invoked as one possible explanation for disky pseudo-bulges \citep{Kormendy2004,Laurikainen2007,Scannapieco2010} and the resulting deviations in the $M-\sigma$ relation \citep{Hu2008,Graham2008}.

Flyby induced impulses have also been suggested as mechanisms for forming kinematically decoupled cores (KDCs) \citep{DeRijcke2004}, tidal tails \citep{D'Onghia2010} and spiral arms \citep{Tutukov2006}, evolution from spiral to S0 Hubble type \citep{Bekki2011}, disk warping \citep{Dubinski2009}, and even increased lithium abundance due to tidal cosmic rays \citep{Prodanovic2012}. Furthermore, ram pressure and tidal stripping during a flyby can quench star formation and lower a galaxy's mass-to-light (M/L) ratio. The result is a population of red galaxies obeying the halo occupation distribution for satellites, but classified as central galaxies \citep{Bahe2013,WTCB13}. 

Our aim is to investigate the ability of flybys to trigger bar formation. \S\ref{S_methods} discusses simulations and analysis techniques, \S\ref{S_results} reports our findings, \S\ref{S_discuss} discusses the results and their role in a larger context, and \S\ref{S_summary} summarizes our findings and outlines future research.

\section{Methods}\label{S_methods}
To assess the ability of flybys to excite bars, we launch three planar interactions -- an equal mass prograde, an equal mass retrograde, and a minor prograde encounter, where the primary and intruder mass ratio is 10:1.

\subsection{Galaxies}\label{SS_galaxies}
We constructed two-component galaxy models with an exponential stellar disk and Hernquist dark matter halo \citep{Hernquist1990} that is $\sim30$ times more massive than the disk. One galaxy was Milky Way (MW)-sized, while the other is 10\% as massive. See Table \ref{tab:components} for a complete description of the galaxy parameters.

We exclude a bulge from the models so any $m=2$ perturbation has maximal impact on the disk. Bulges are thought to stabilize disks against bar formation \citep{Shen2004,Athanassoula2005} and are thus omitted from this initial simulation suite. 

The gravitational softening is set by the inter-particle separation at the disk scale length. The halo softening length is scaled from this by $\sqrt{m_{\rm p,halo}/m_{\rm p, disk}}$.

The initial conditions were evolved in isolation for 3\unit{Gyr} using {\sc Gadget-2} \citep{Springel2001,Springel2005b} to ensure that the system was close to dynamical equilibrium and assess the perturbations induced by secular evolution. This is especially important because artificial halo truncation drives minor evolution of the system.

\subsection{Interactions}\label{SS_interactions}
We selected planar orbits with impact parameters equal to 10\% of $R_{\rm vir,1}+R_{\rm vir,2}$. To explore the effect of mass ratio and inclination angle, we model 1:1 prograde, 10:1 prograde, and 1:1 retrograde interactions. Interactions start with a separation of $R_{\rm vir,1}+R_{\rm vir,2}$ and velocity such that $v_{\rm peri}=2v_{\rm circ}$. Being at the energetic boundary of a merger and flyby, we consider the dynamical excitations produced here to be a lower limit for the typical flyby.

Although planar interactions are cosmologically rare, planar prograde and retrograde interactions maximally torque the disk. Since we are interested in maximizing the impact of the interactions, intermediate inclination angles are left for future study.

Interactions were evolved for a minimum of 5\unit{Gyr} using {\sc Gadget-2} \citep{Springel2001,Springel2005b} with snapshots every 0.05\unit{Gyr}, capturing pericenter and several dynamical times afterward to study the persistence of any perturbations.

\subsection{Analysis}\label{SS_methanalysis}

\subsubsection{Spherical Harmonics}\label{SSS_methbarstrength}
To assess bar strength, disk potentials are decomposed in spherical harmonics using a self-consistent field (SCF) approach \citep[cf.][]{Hernquist1995a}. As bars are traced by the $m=2$ part of the potential ($\Phi_{m=2}$), the amplitude compared to the total potential ($\Phi_{\rm tot}$) defines an $m=2$ amplitude ($A_{m=2}=\Phi_{m=2}/\Phi_{\rm tot}$) at any position.  By projecting azimuthally averaged $A_{m=2}$ onto polar coordinates in the plane of the disk, we can identify borders of positive $m=2$ regions signifying a bar, and define an overall bar amplitude as the maximum value of $A_{m=2}$ within these borders. 

We define a bar as a region of positive $A_{m=2}$ with a constant position angle that does not change by greater than $10^{\circ}$ along its length with amplitude $A_{m=2}\gtrsim0.04$ \citep[cf.][]{Athanassoula2013}. The radial extent of this region is a proxy for bar length ($l_{m=2}$). 

\subsubsection{Ellipse Fitting}\label{SSS_methbarshape}

As a cross-check, we use an observationally-based technique which fits ellipses to isophotes; the change in ellipses with radius defines the bar \citep[See][]{Jedrzejewski1987}.
Here we fit ellipses to a proxy for surface brightness, the projected mass density of the face-on disk ($I$). For a given semi-major axis ($a$), an ellipse is characterized by a center ($X0$,$Y0$), ellipticity ($\varepsilon$), and position angle ($\phi$) and is fit to the projected mass density as follows:
\begin{enumerate}
\item Find azimuthally averaged intensity at $nE$ eccentric anomalies ($E$) along a trial ellipse and calculate residuals from the mean intensity along the ellipse.
\item Find $dI/da$ via finite differencing between neighboring ellipses within $derwid$ of the trial ellipse's semi-major axis, but with the same center, ellipticity, and position angle.
\item Express residuals as a 2$^{\rm nd}$ order Fourier expansion in $E$ to obtain 4 harmonics ($A_{1}$, $B_{1}$, $A_{2}$, $B_{2}$) and the root-mean-square error ($rms$).
\item Ellipse parameters are accepted if: 
\begin{itemize}
\item a maximum number of iterations is exceeded ($maxIter$),
\item $dI/da$ is less than $dertol*rms$
\item the largest harmonic is less than $errtol*rms$ and a minimum number of iterations ($minIter$) has been met
\end{itemize}
Otherwise....
\item Apply the largest harmonic as a correction factor to the corresponding ellipse parameter such that the harmonic is reduced to zero \citep[See][]{Jedrzejewski1987}, reducing the correction factor by 1\% with each iteration to ensure convergence.
\item The process is then repeated for a new trial ellipse with a different $a$.
\end{enumerate}
We set $nE=100$, $derwid=0.05$, $dertol=0.5$, $errtol=0.04$, $minIter=8$, and $maxIter=20$ and verified that results do not change for a 10\% increase/decrease in each parameter. When iterations are ceased but error tolerance is not met, the set of parameters corresponding to the iteration with the smallest harmonic is selected.


For each snapshot, we extract bar parameters from the ellipse fits at 20 radii \citep[See][]{Marinova2007}. With this technique, we define a bar as a region of nearly constant angle ($<10^{\circ}$ along its length) with an ellipticity that rises with increasing radius to $\gtrsim0.25$ and then decreases by $\gtrsim0.1$ at the bar edge ($l_{\rm ell}$). Bars smaller than three image bins are excluded and maximum ellipticity ($e_{\rm max}$) is used as a proxy for bar strength.

\subsubsection{Impact of Spiral Pattern}\label{SS_spiral}
Regardless of technique, coincident spiral patterns make identifying and characterizing bars difficult. For one, both bars and two-armed spiral patterns reflect $m=2$ modes.
Since bars and spirals dominate in different regions, distinguishing the two contributions is relatively simple when they are misaligned. However, when the bar and spiral pattern speeds differ, periodic alignment prevents clean measurement of the bar contribution. This is seen as periodic variation in the bar amplitude due to confusion with the overlapping spiral pattern (See Figures \ref{fig:prograde1} \& \ref{fig:prograde10_2}). Therefore, while large scale trends can be trusted, short duration features cannot.

Ellipse fits are also affected. While fits to strong bars are generally well-behaved and converge, harmonic expansion along an ellipse intercepting a spiral pattern will exhibit additional power in the harmonic corresponding to the number of arms. For example, an ellipse intercepting a two-armed spiral will have larger $A_{2}$ or $B_{2}$ harmonics and may not converge on a position angle or ellipticity if the spiral feature is not subtracted. However, since we are interested in fitting the inner disk where a bar would dominate, we ignore transient effects driven by the spiral.

\section{Results}\label{S_results}

\subsection{1:1 Prograde}\label{SS_prograde1}
\begin{figure*}
	\snaprow{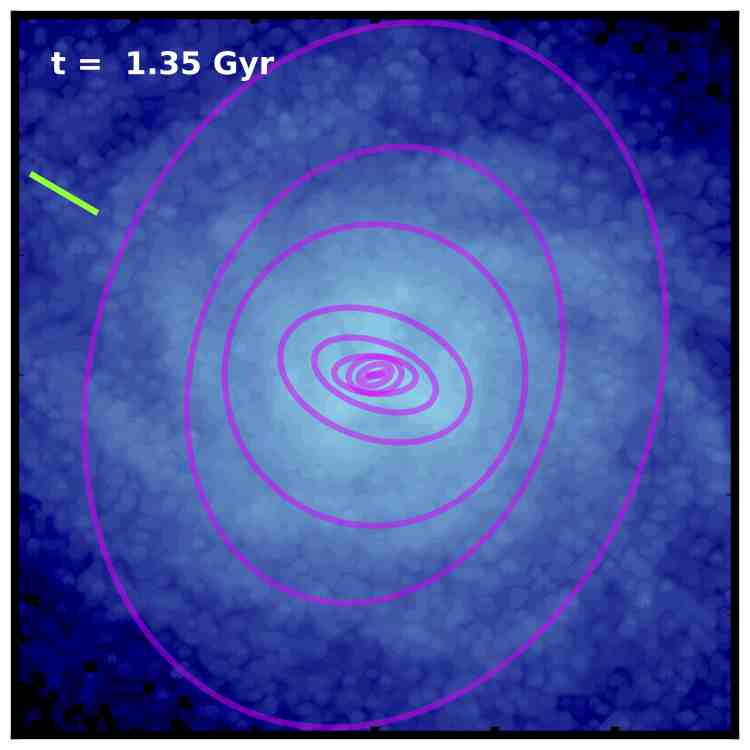}{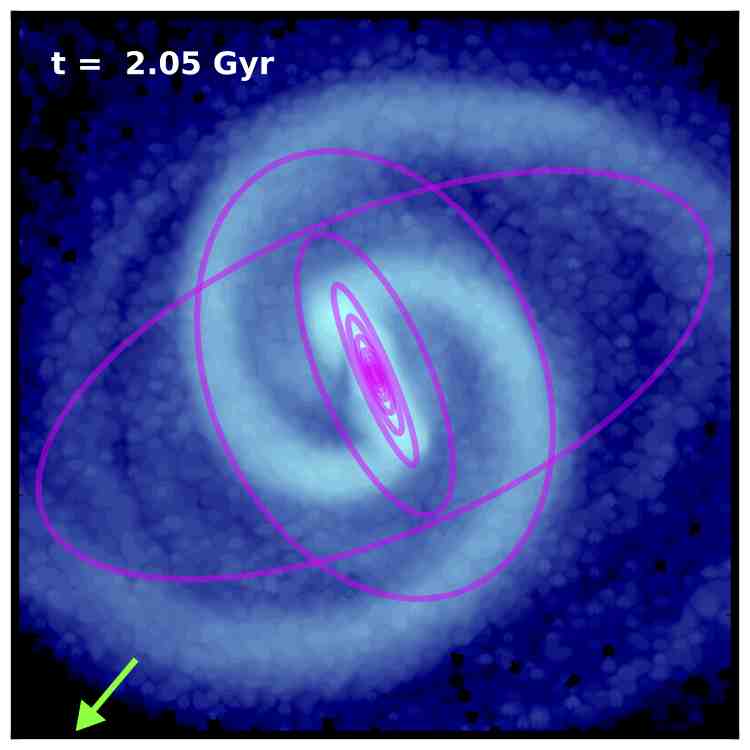}{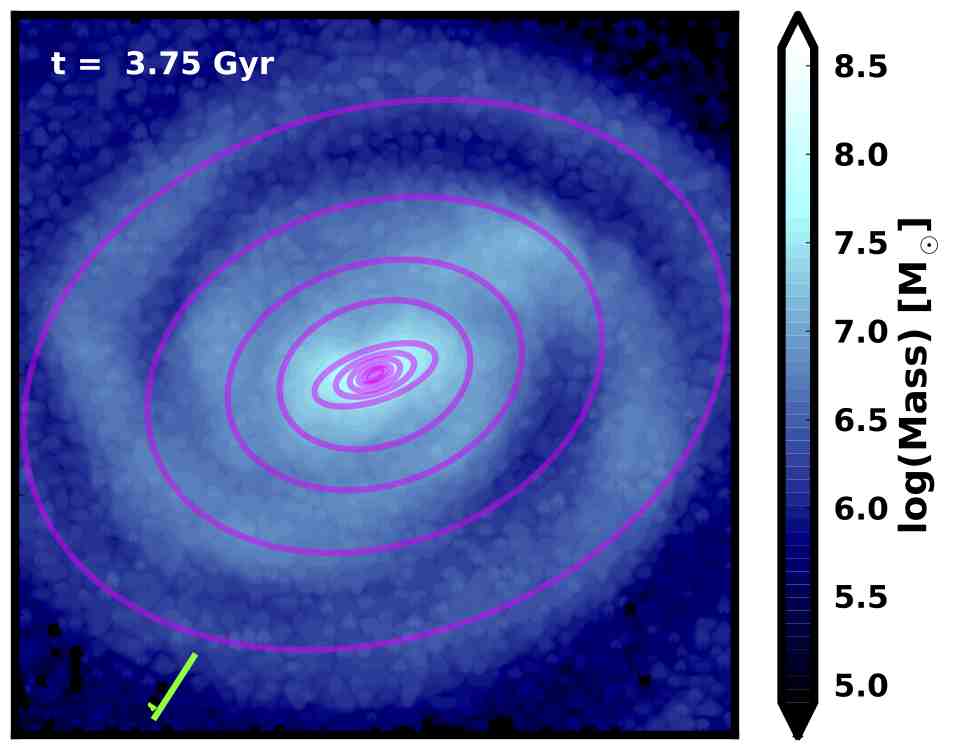}
	\snaprow{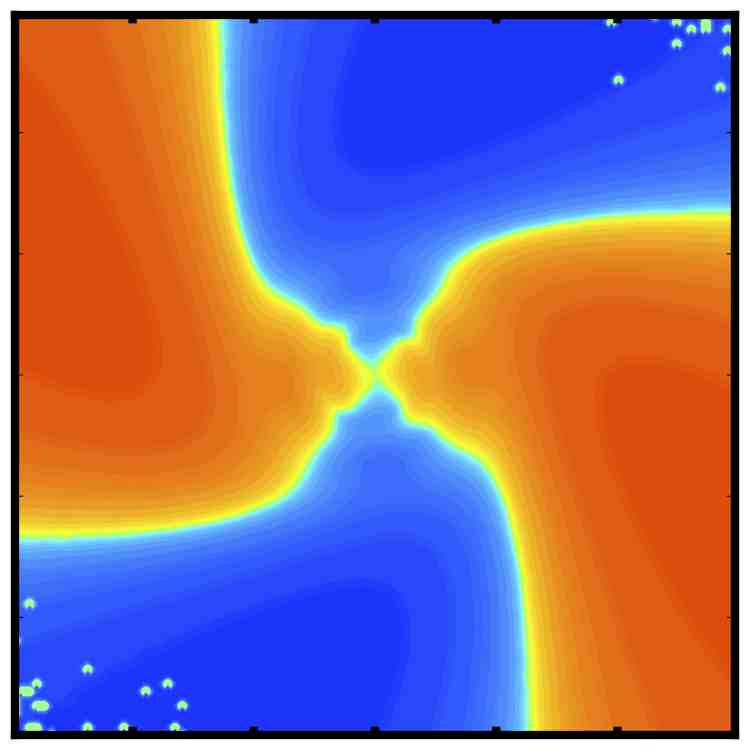}{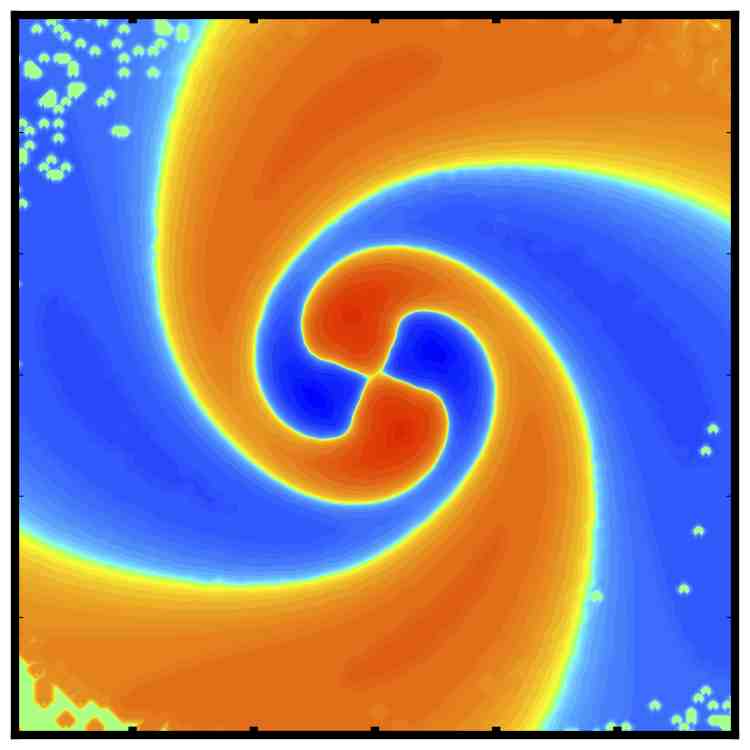}{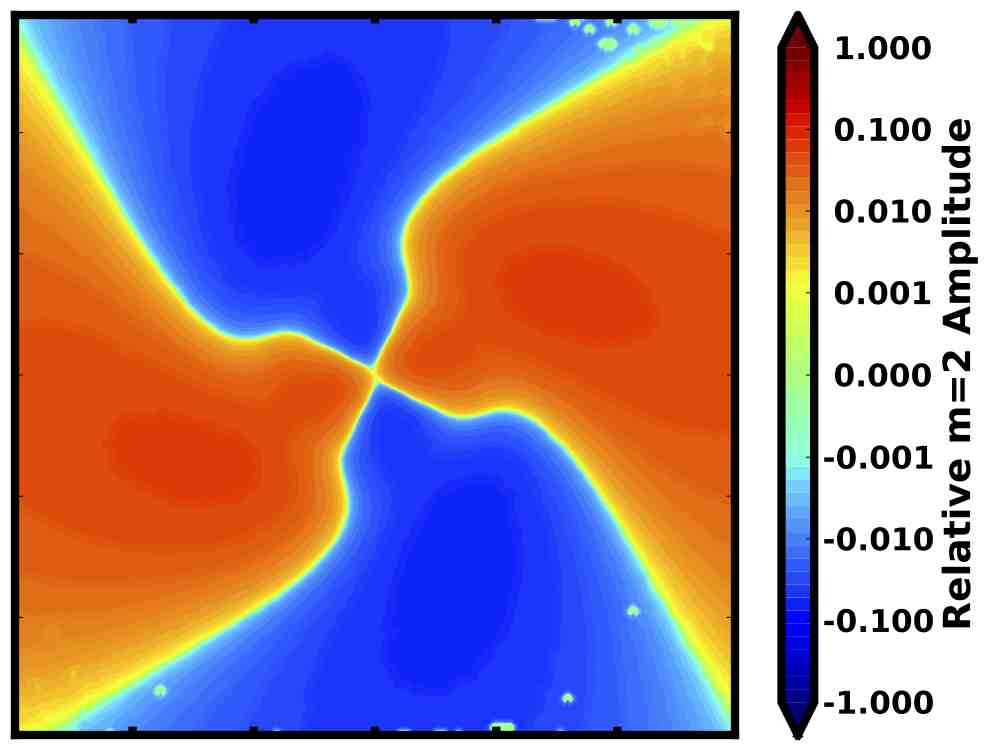}
	\evolrow{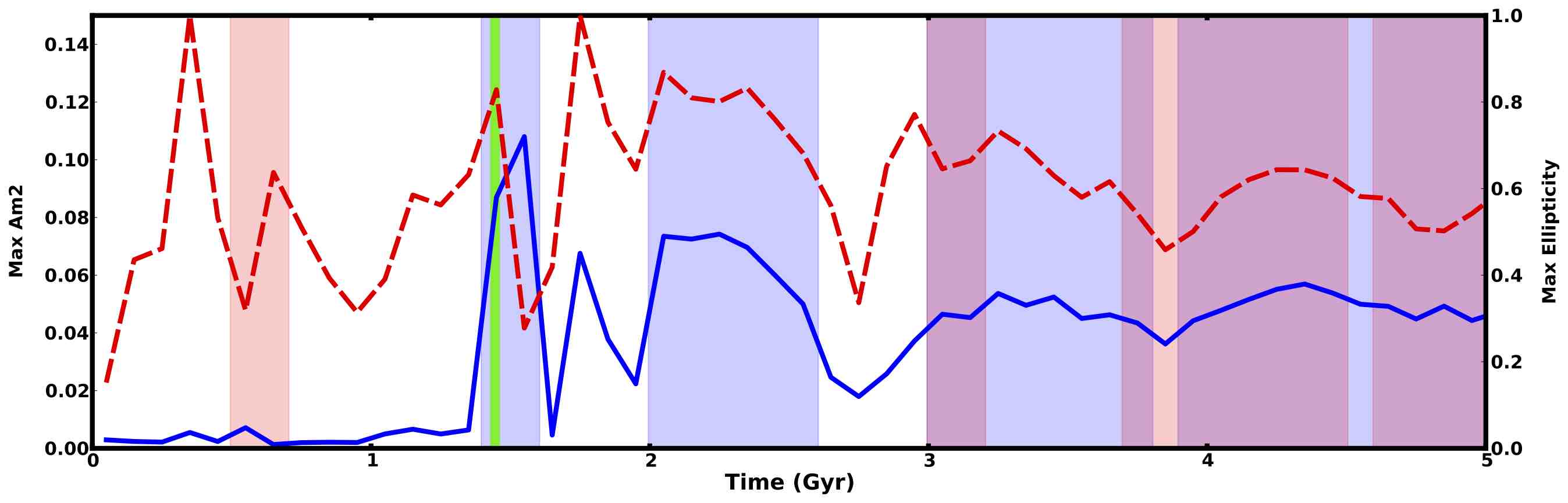}
    \caption{1:1 Prograde Primary. Row 1: Face-on projected mass distributions of the inner 30\unit{kpc} of the disk for three times during the interaction. Ellipse fits are in magenta, times are printed in each column, and the direction to the secondary is marked with a green arrow. Row 2: Projected $m=2$ amplitude. Row 3: Time dependence of $A_{m=2}$ (left axis, blue solid line) and $e_{\rm max}$ (right axis, red dashed line). The blue and red regions meet requirements for positively identifying a bar using SCF and ellipse methods respectively. The solid green vertical line marks pericenter. \label{fig:prograde1}} 
\end{figure*}

\begin{figure*}[ht]
	\snaprow{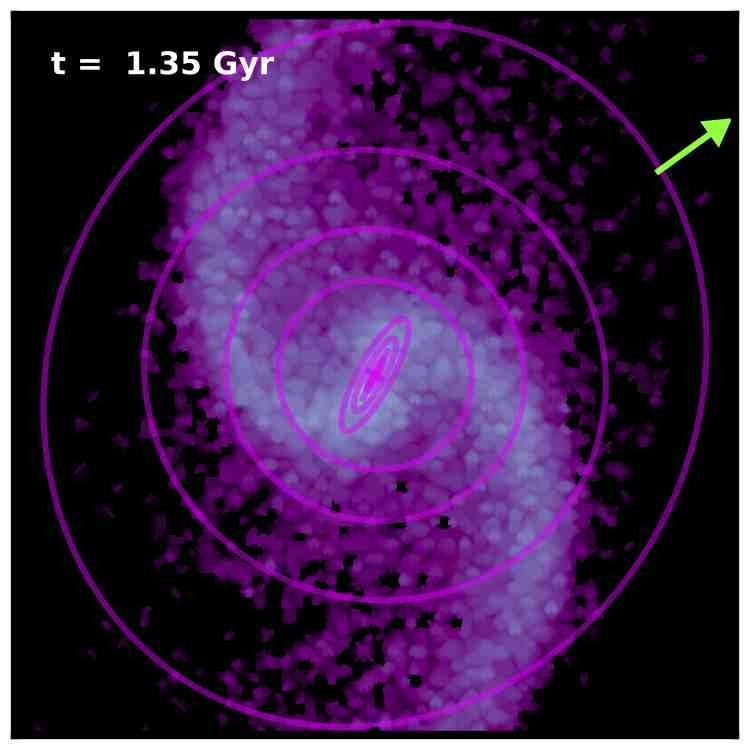}{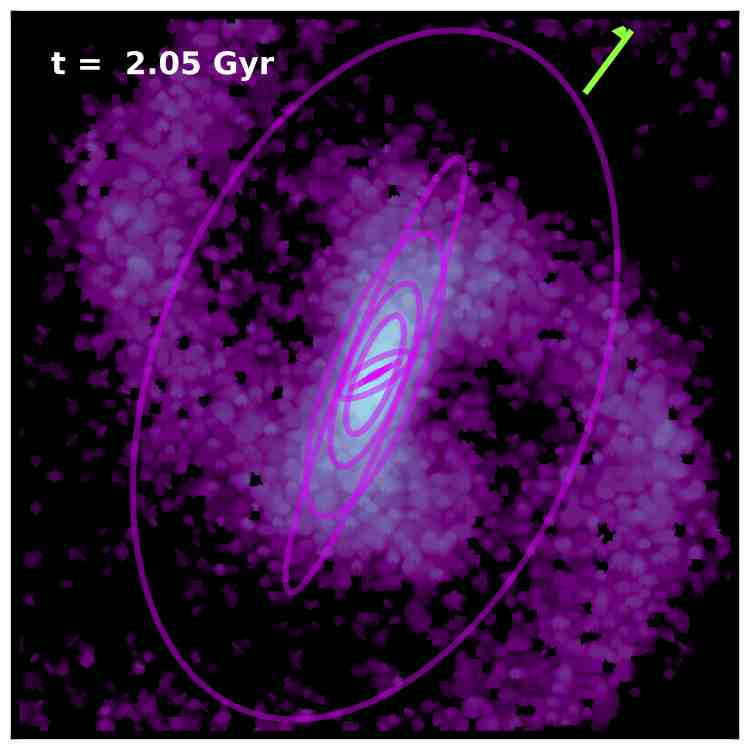}{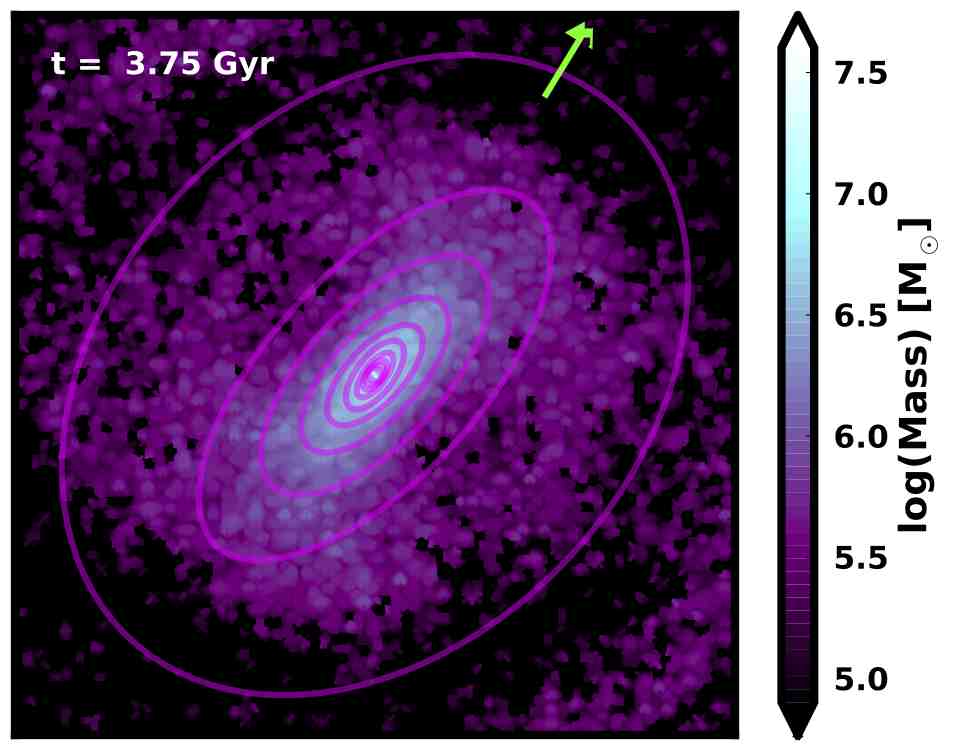}
	\snaprow{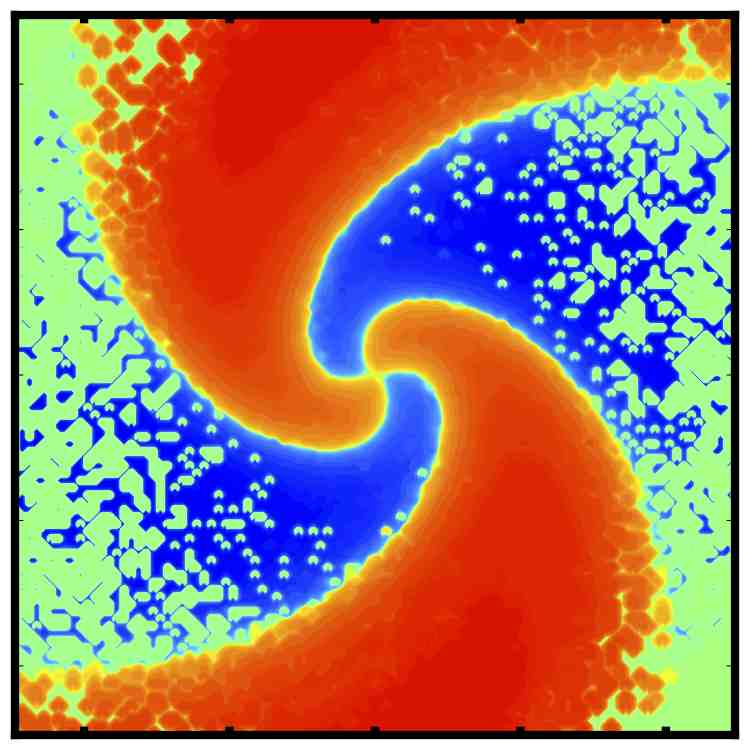}{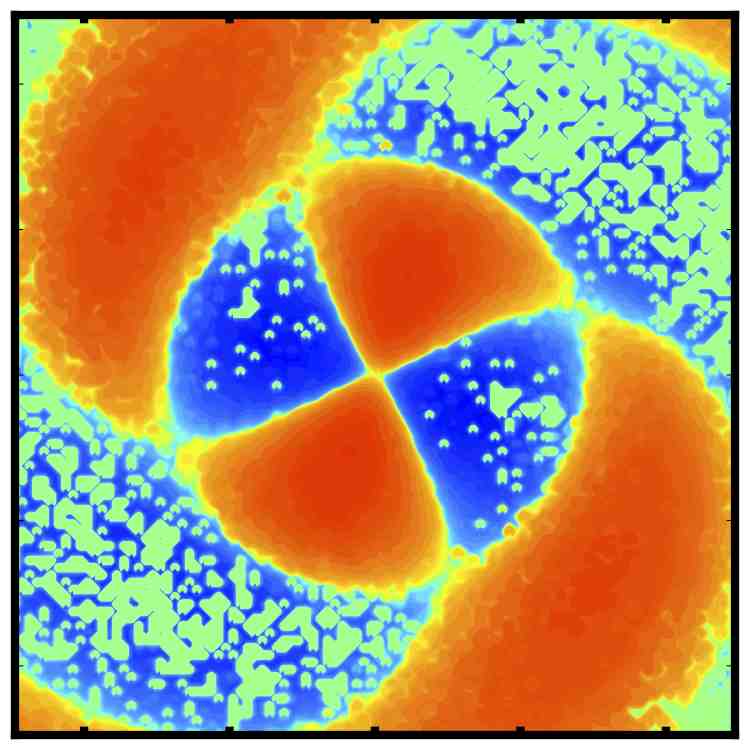}{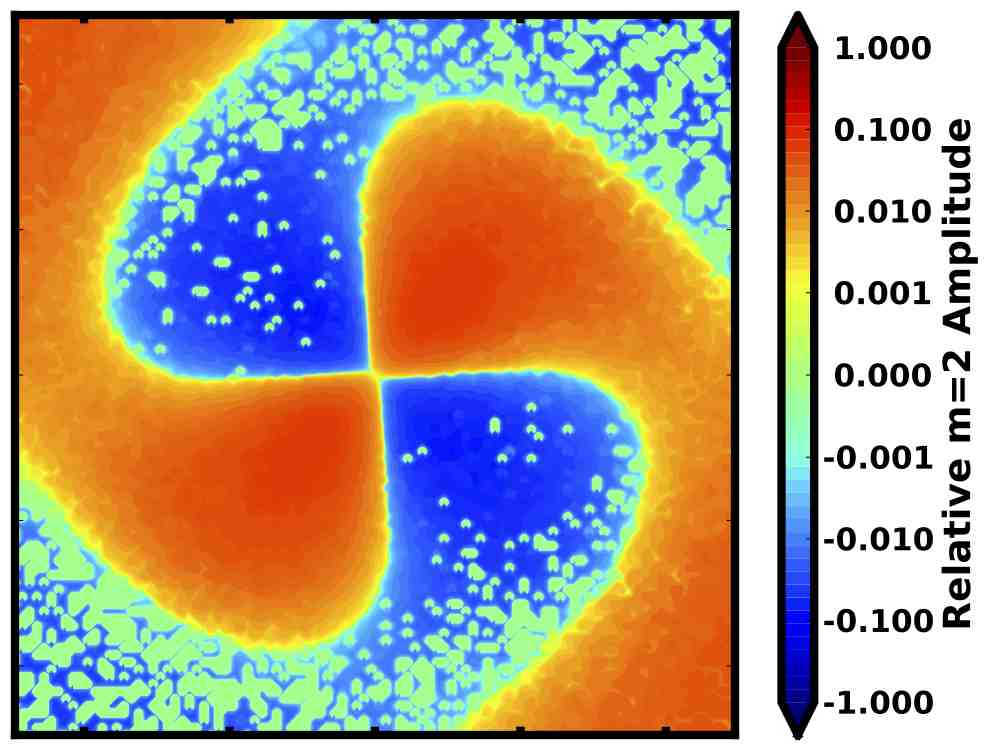}
	\evolrow{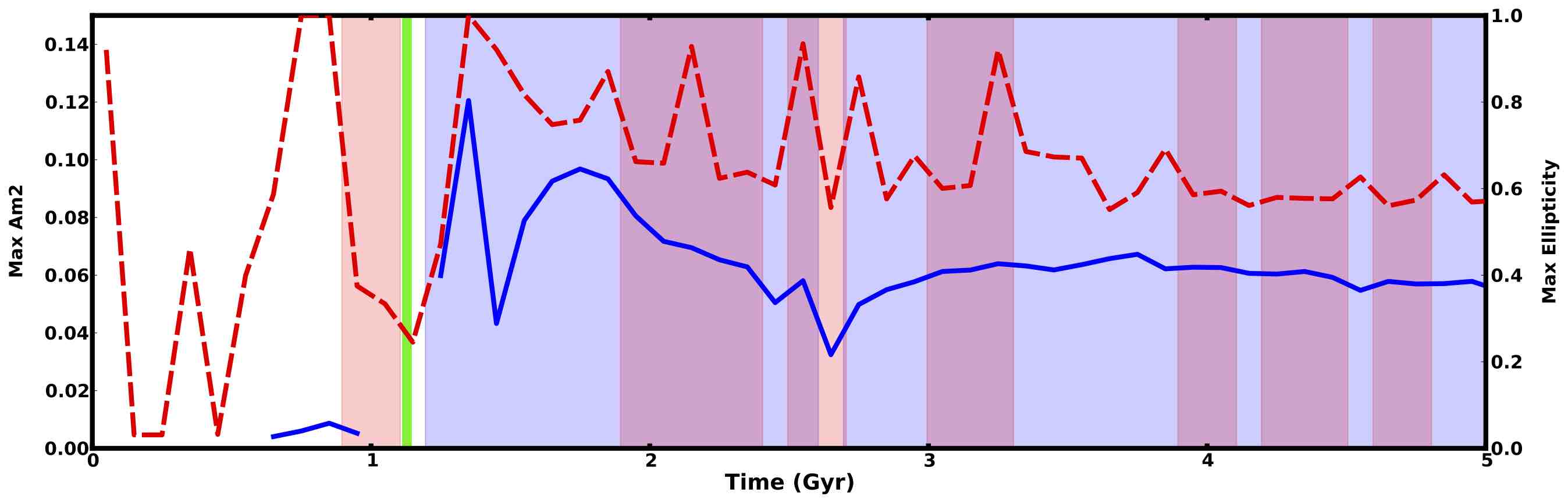}
  \caption{10:1 Prograde Secondary. Same as Figure \ref{fig:prograde1}, but zoomed in on the central 10\unit{kpc}. \label{fig:prograde10_2}}
\end{figure*}

Figure \ref{fig:prograde1} shows the primary at various stages throughout the 1:1 prograde interaction. Both galaxies experience significant perturbation and both analysis techniques identify a bar (bottom row, Figure \ref{fig:prograde1}).  


Prior to pericenter, neither method identifies a bar. $A_{m=2}$ remains below 0.01 and $e_{\rm max}$ varies widely. The disk features a transient three-armed spiral within $\sim4$\unit{kpc} that is also present when evolved in isolation. In isolation, this feature is dominated by fluctuating $m=1$ and $m=3$ modes that never grow above 2\% of the potential. These modes also appear intermittently throughout the prograde interaction, however the interaction induces a much stronger $m=2$ mode reaching $>10\%$ of the potential. Given that these odd modes are transient and constitute a small fraction of the potential, we believe that our conclusions on the ability of flybys to form bars are robust.

$A_{m=2}$ rises to a peak of 0.11 just 0.1\unit{Gyr} after pericenter with intermittent bar identification. Although our ellipse fitting bar criteria is not satisfied during this period, we begin to see $e_{\rm max}$ trace $\sim$1\unit{Gyr} periodic variations in $A_{m=2}$.

At $\sim3$\unit{Gyr}, $A_{m=2}$ and $e_{\rm max}$ level out to $0.05$ and $0.6$ respectively, corroborating the presence of a bar rotating at 8.9\unit{rad/Gyr}. While the more conservative $m=2$ method indicates a bar length of $l_{\rm m=2}=3.8$\unit{kpc}, the ellipse fitting technique points to a longer bar with $l_{\rm ell}=8.1$\unit{kpc}. 

In addition to a bar, both galaxies exhibit a two-armed spiral that winds and begins to dissipate $\sim5.5$\unit{Gyr} after pericenter. As discussed in \S\ref{SS_spiral}, the presence of both a bar and spiral pattern causes periodic variation in $A_{m=2}$ every $\sim$1\unit{Gyr} as the bar and spiral align. 

\subsection{10:1 Prograde}\label{SS_prograde10}
\subsubsection{Primary}
\begin{figure*}[ht]
	\snaprow{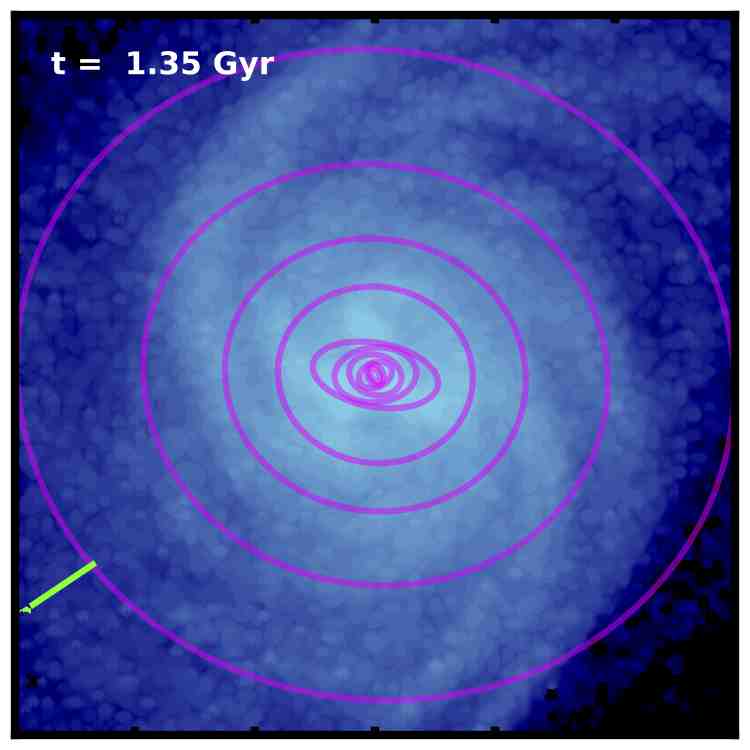}{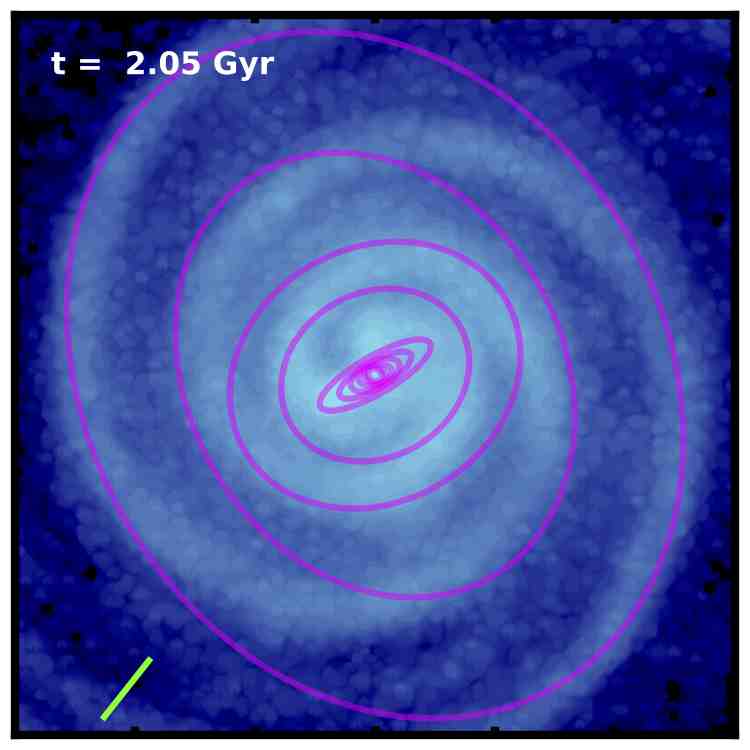}{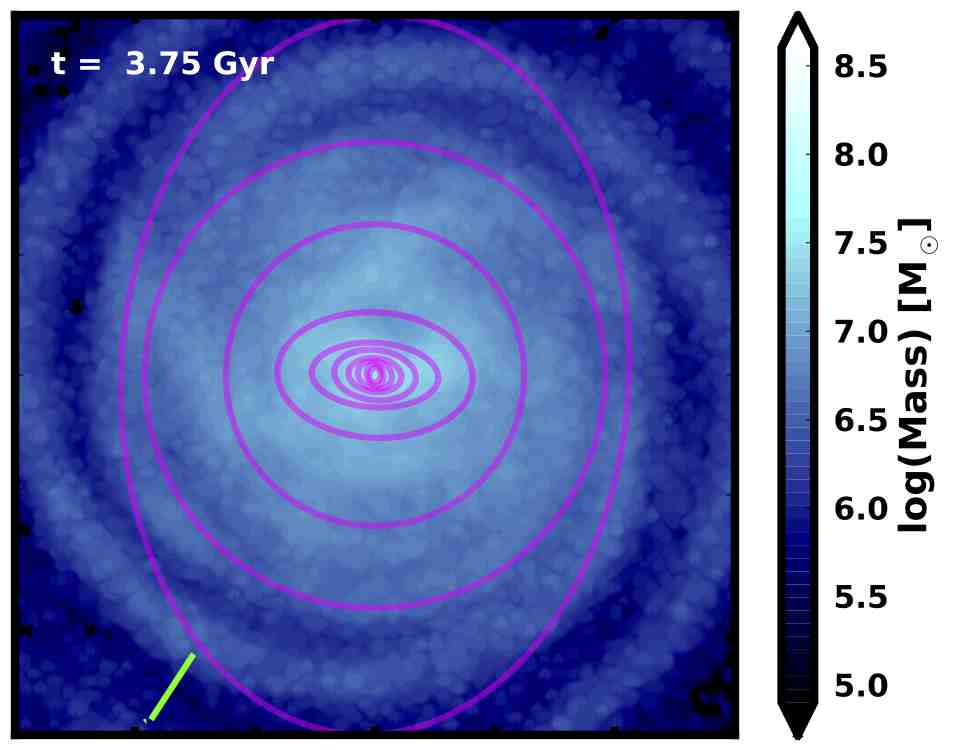}
	\snaprow{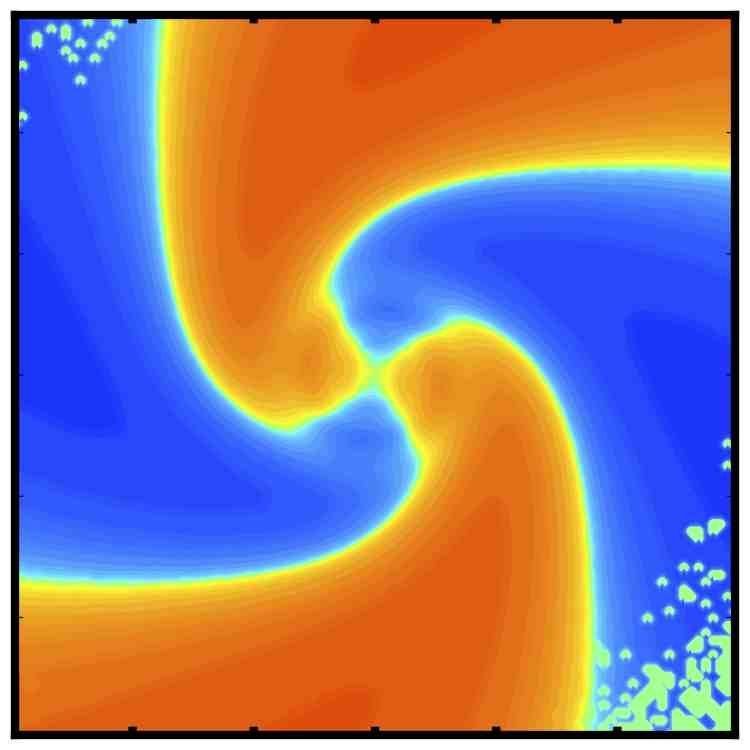}{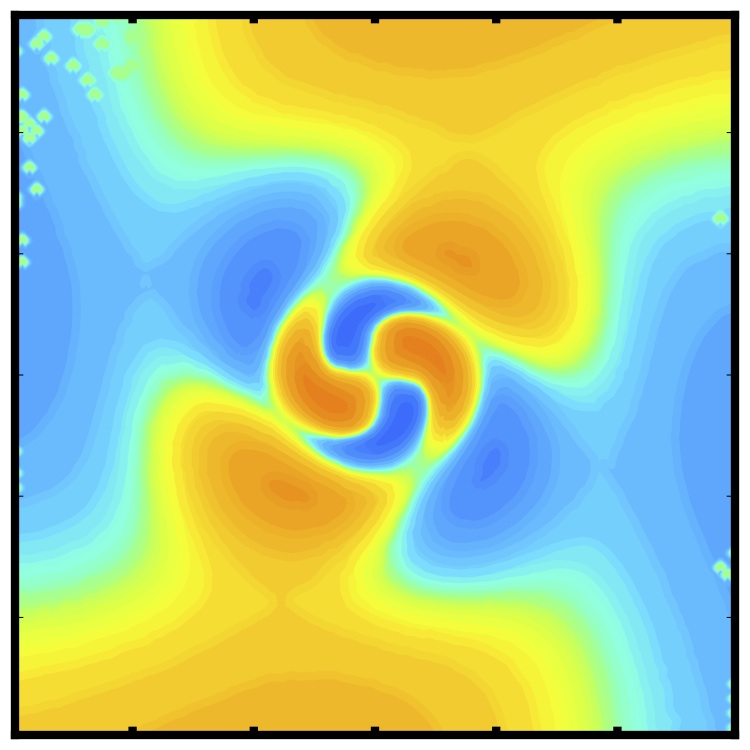}{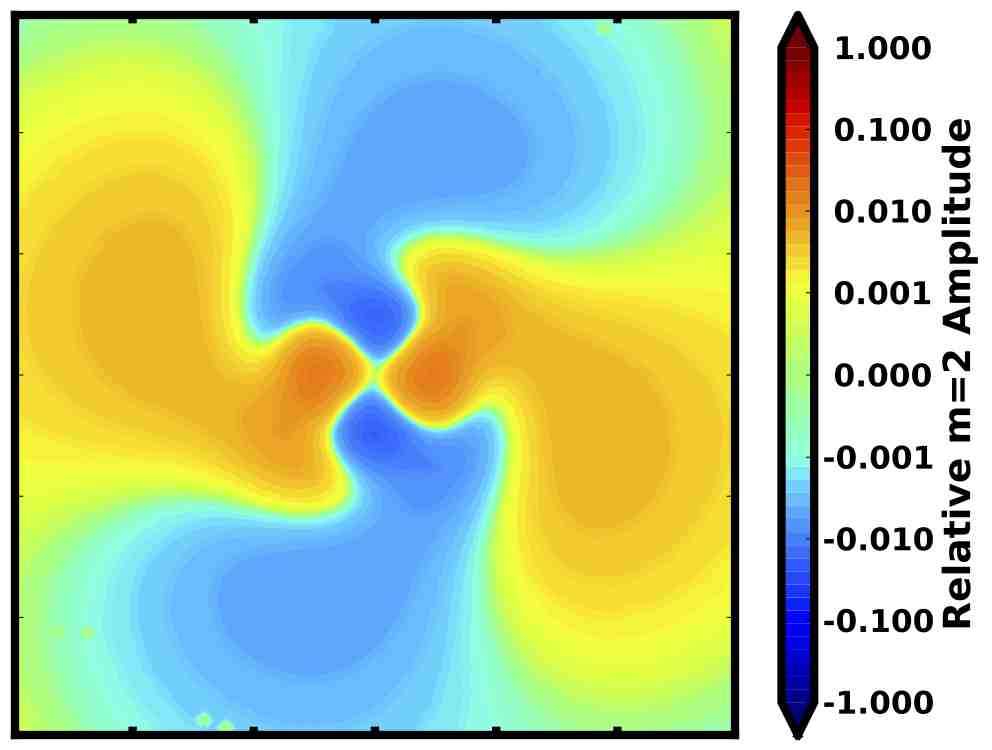}
	\evolrow{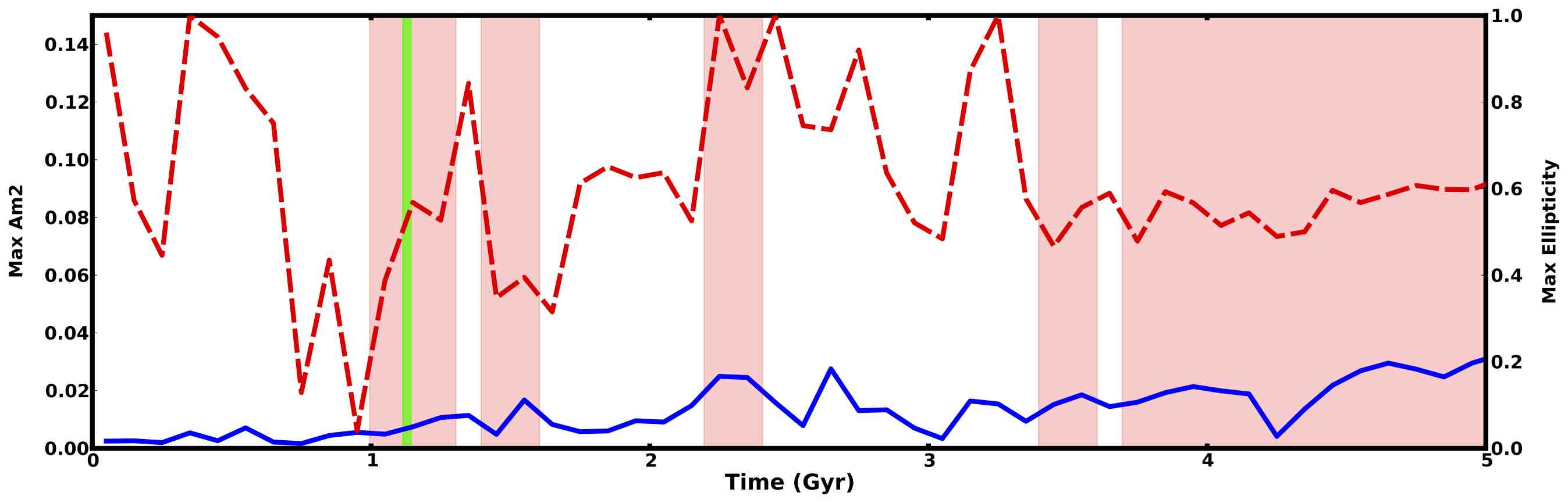}
  \caption{10:1 Prograde Primary. Same as Figure \ref{fig:prograde1}. \label{fig:prograde10}}  
\end{figure*}

The 10:1 prograde interaction has a subtle impact on the primary (Figure \ref{fig:prograde10}). Although a bar is absent following pericenter at 1.13\unit{Gyr}, the disk features a both a two-armed spiral ($m=2$) and a transient three-armed feature ($m=3$) in its central $\sim4$\unit{kpc}. When the two patterns align, the stronger two-armed spiral destroys one arm in the three-armed feature. This destruction shifts $m=3$ power to the bar mode and causes poor ellipse fits intersecting the missing arm. This results in a periodic increase in $A_{m=2}$ that is somewhat echoed in $e_{\rm max}$ (Figure \ref{fig:prograde10}, bottom panel). 

While the $A_{m=2}$ parameter peaks and then levels out during the 1:1 prograde interaction, $A_{m=2}$ increases throughout the entire 10:1 interaction to $\sim0.03$ ($l_{m=2}=3.1$\unit{kpc}) about 3.3\unit{Gyr} after pericenter, half the final value in the 1:1 interaction and below the level signifying a bar. While the spiral arms cause ellipse fits to be erratic during the first part of the simulation, $e_{\rm max}$ levels out to 0.58 around 3.5\unit{Gyr} after pericenter as the two-armed spiral winds and disappears. By $\sim4$\unit{Gyr} after pericenter, the destruction of the third arm is enough that the ellipse fits identify a $l_{\rm ell}=5.2$\unit{kpc} bar rotating at 9.4\unit{rad/Gyr}. However, due to the lack of evidence in $A_{m=2}$, this is not considered a reliable identification.

\subsubsection{Secondary}
Unlike the primary, the 10:1 secondary disk is strongly perturbed (Figure \ref{fig:prograde10_2}). A two-armed spiral forms just 0.2\unit{Gyr} after pericenter (at 1.13\unit{Gyr}) with $A_{m=2}=0.12$. As the two galaxies separate, a bar grows to occupy the majority of the disk. The $A_{m=2}$ levels out to 0.06 with a size of $l_{m=2}=2.4$\unit{kpc} and $e_{\rm max}$ levels out to 0.6 with $l_{\rm ell}=2.7$\unit{kpc}. Both analyses methods intermittently identify a bar from pericenter on. By $\sim4$\unit{Gyr} after pericenter, the disk is dominated by a bar rotating at 7\unit{rad/Gyr} and the spiral arms are weak and transient. 

Under-sampling of the surface density introduces noise into the ellipse fits that translates to a large variation in $e_{\rm max}$. While $A_{m=2}$ is also affected to some degree (responsible for the break at pericenter), the strongest parts of $m=2$ mode coincide with high density, and thus highly sampled, regions.

\subsection{1:1 Retrograde}\label{S_retrograde1}
\begin{figure*}[ht]
	\snaprow{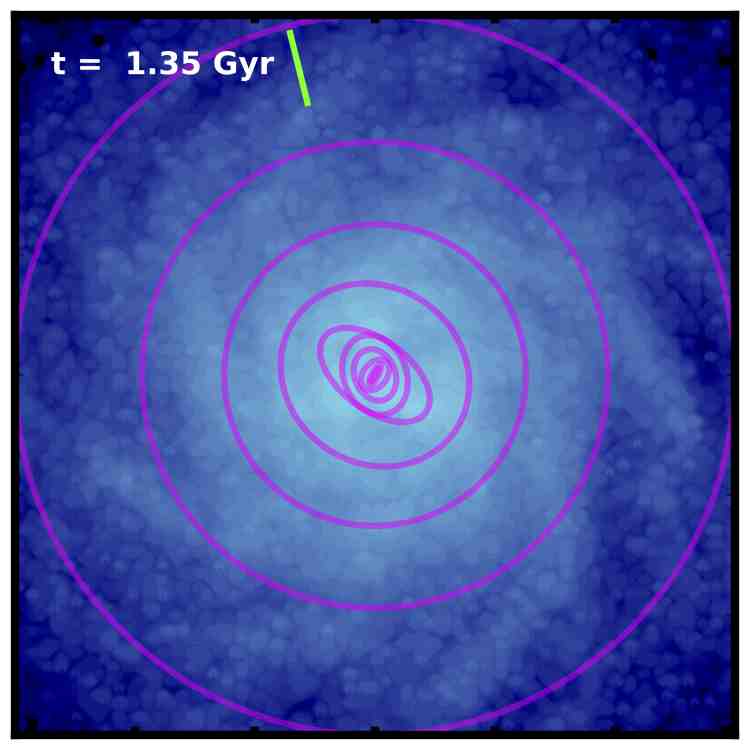}{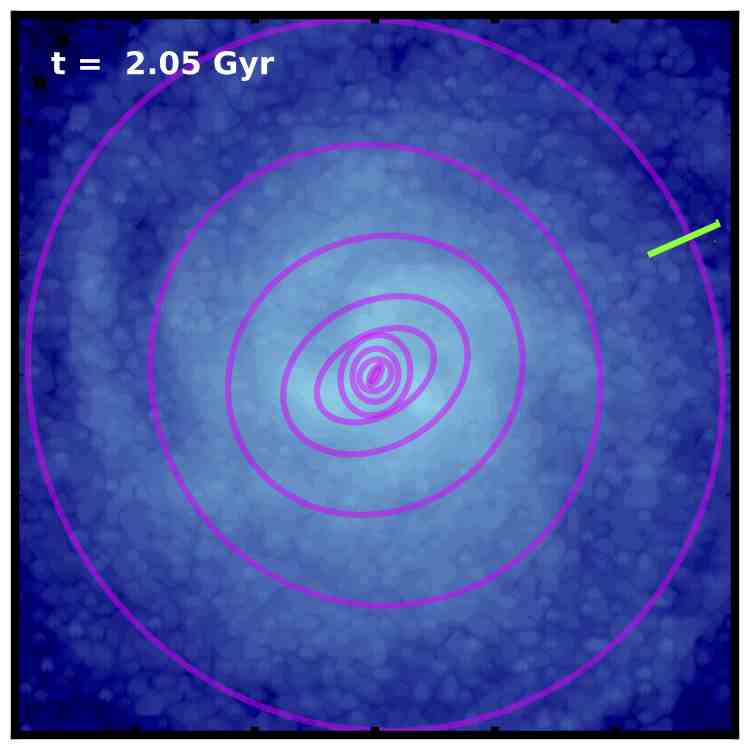}{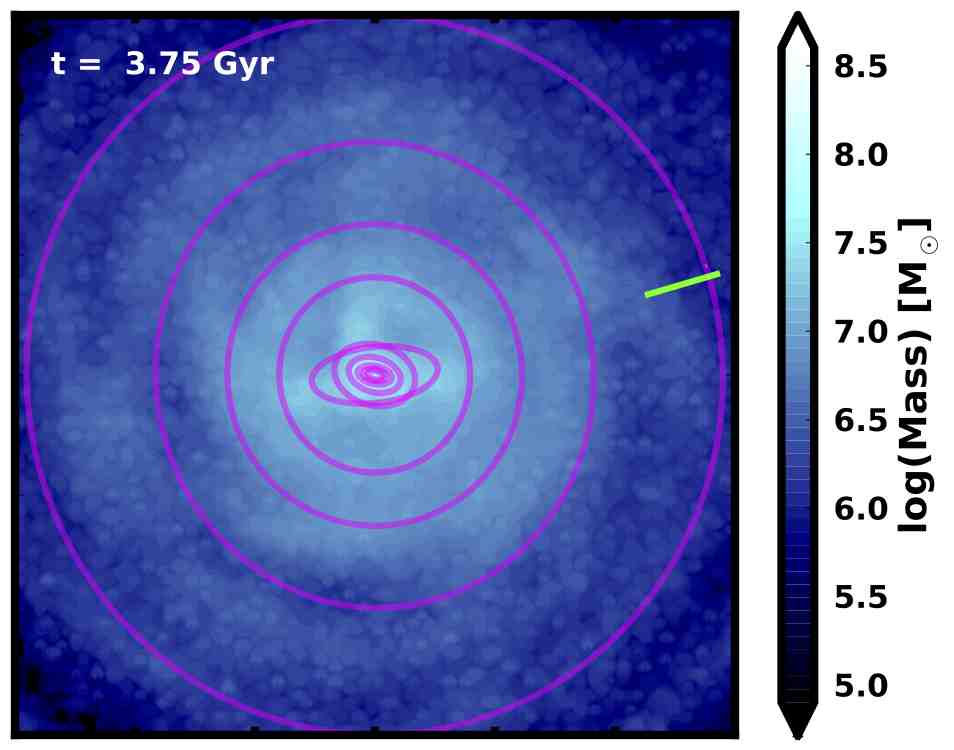}
	\snaprow{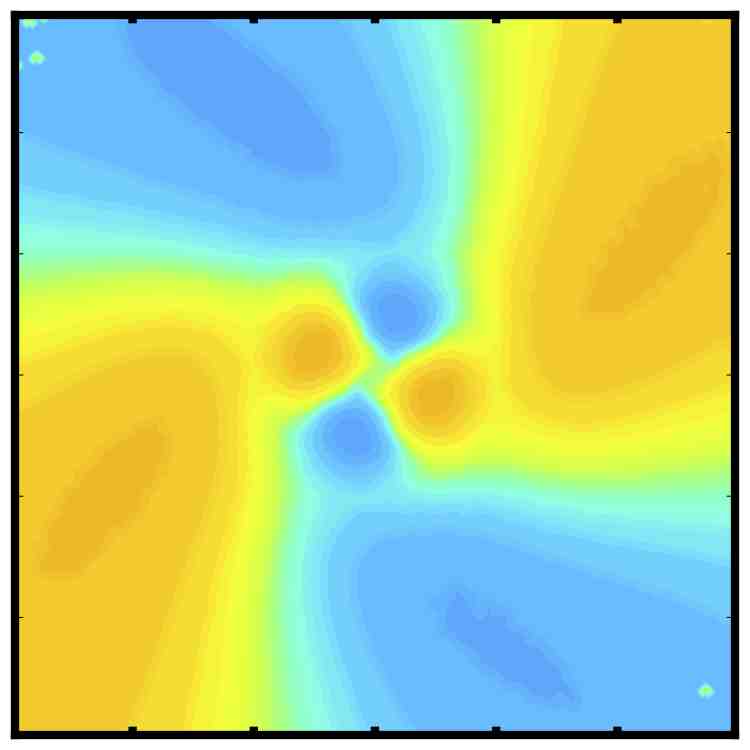}{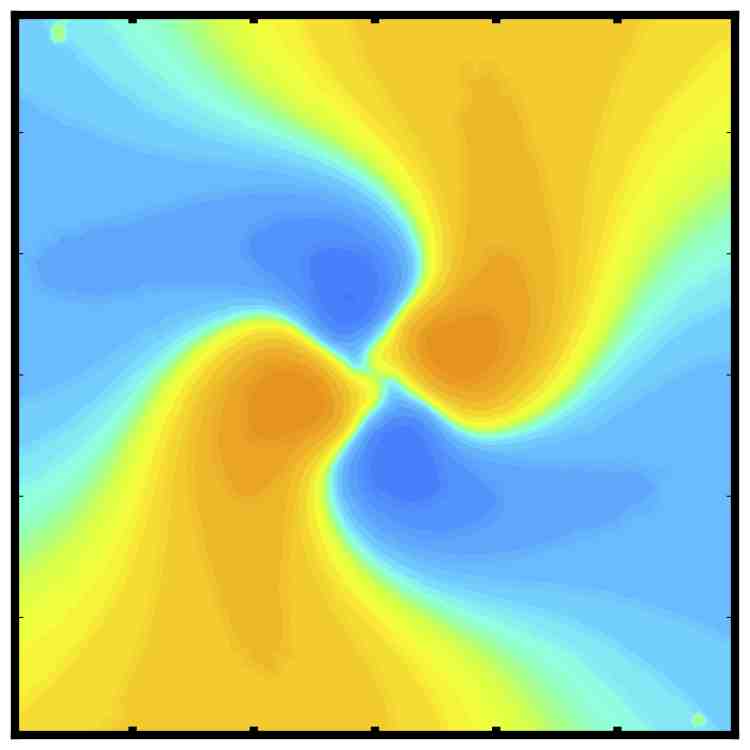}{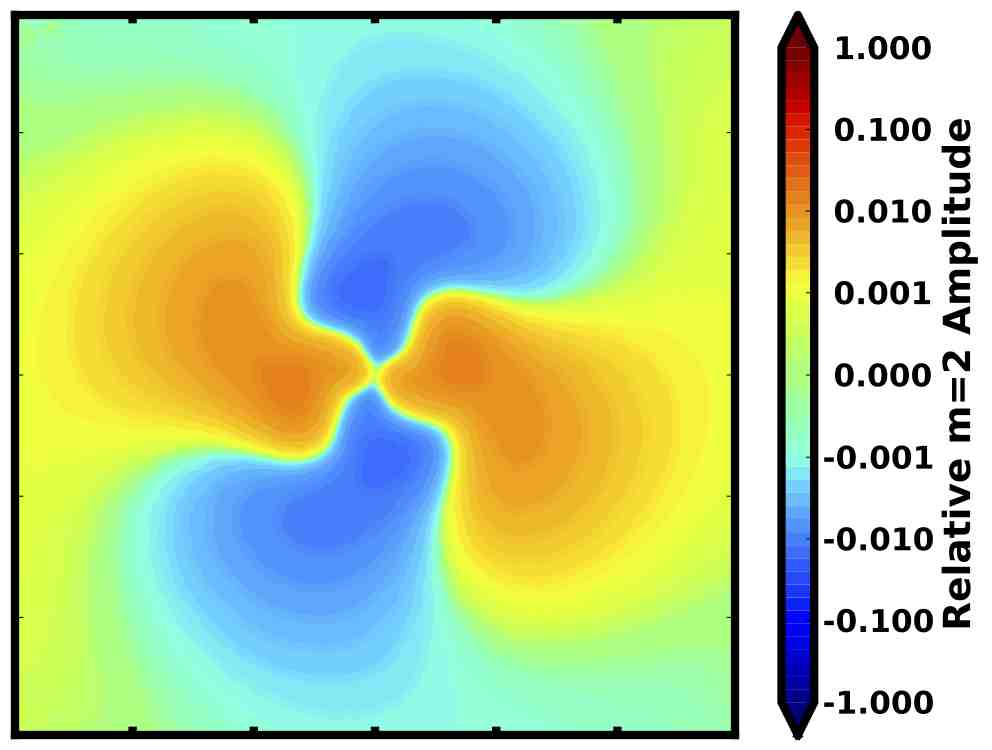}
	\evolrow{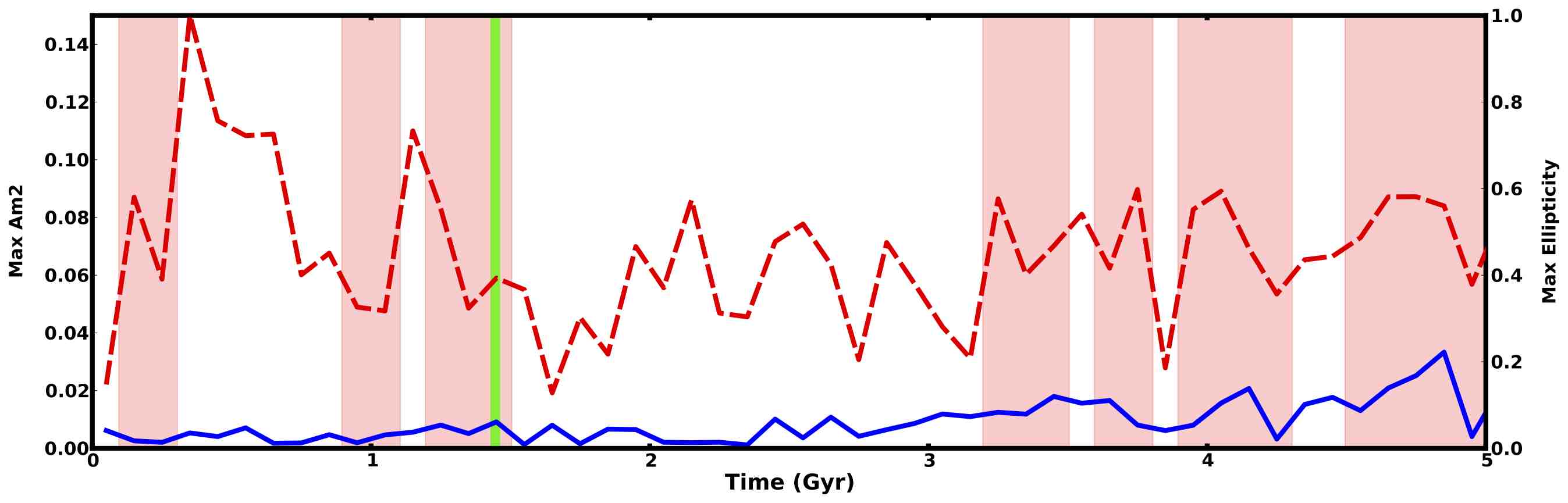}
  \caption{1:1 Retrograde Primary. Same as Figure \ref{fig:prograde1}. \label{fig:retrograde1}}  
\end{figure*}

Neither galaxy is strongly perturbed by the 1:1 retrograde interaction (Figure \ref{fig:retrograde1}) and only a transient three-armed spiral is discernible. However, as this is also seen in isolation, it is likely unrelated to the interaction. 

Although $A_{m=2}$ and $e_{\rm max}$ increase throughout the simulation, $A_{m=2}$ never exceeds 0.02 and this is reflected in the absence of a bar or two-armed spiral. Ellipse fits identify a 5\unit{kpc} bar intermittently toward the end of the simulation. However, as in the primary of the 10:1 prograde interaction, this is likely due to fitting of two arms from the inner 3 armed spiral and is not corroborated by the $m=2$ analysis.

\section{Discussion and Potential Implications}\label{S_discuss}

Interactions are expected to have a smaller impact on the primary and larger impact on the secondary as mass discrepancy increases. Because the 1:1 prograde interaction produced a primary bar and the 10:1 did not, the maximum mass ratio to produce a primary bar for these orbital parameters is somewhere in-between and additional simulations are required to pin it down. However, both the 1:1 and 10:1 prograde flybys induced bar formation in the secondary, implying that planar flybys with a smaller impact parameters or orbital eccentricities should also form secondary bar as well.

As expected, the retrograde interaction had negligible impact on either galaxy. While planar orbits were selected to maximize the impact, they are cosmologically rare. As inclination angle increases above the plane, an interaction's ability to induce a bar decreases, but the angle beyond which bar formation no longer occurs will depend on other properties like mass ratio and eccentricity. Quantifying this dependency requires further simulations and is of interest for future studies.

The formation and persistence of flyby-driven stellar bars has important consequences for the observed bar fraction in seemingly isolated galaxies. In our simulations, bars persist several Gyr after the interaction when the galaxies were separated by Mpc. Thus, many bars observed in isolated galaxies may be the result of flyby interactions rather than secular evolution. Like mergers, flybys are more likely in high density environments \citep{SH12} and flyby-induced bars may enhance the bar fraction in the outskirts of these regions.

Surprisingly, \citet{SH12} found high mass ratio flybys are as common as mergers at low redshift. Since the secondary, `intruder' galaxy will be more strongly affected, flyby-triggered bars could also enhance the bar fraction for low mass galaxies at low redshift. In fact, \citet{Sheth2008} found that, while bar fraction increases with mass at high redshift ($z=0.60-0.64$), the bar fraction appears to be independent of mass at low redshift ($z=0.14-0.37$) -- this may point to an increase of flyby-induced bars in low mass galaxies at the present epoch. 

Resonant interactions transfer energy and angular momentum from the bar to underlying dark matter halo. This may flatten the inner halo density profile \citep[][e.g]{Holley-Bockelmann2005} and even form a bar in the halo itself \citep{Holley-Bockelmann2005,Athanassoula2007}. Bar-induced flattening of the central halo may alleviate some of the discrepancy between observations indicating cores and cusps seen in collisionless simulations \citep{DeBlok2010, WeinbergKatz2002}, but accurately resolving the angular momentum exchange requires fine resolution of the appropriate resonances \citep{Dubinski2009a, Weinberg2007a}. 


Bars also enhance radial migration in disks, causing dynamical heating and chemical mixing \citep[][e.g.]{Minchev2010}. Findings by \citet{Bovy2012} indicate that the MW may not have a distinct thick disk, implicating a secular heating mechanism like bar-induced radial migration \citep{Roskar2013}. Regardless of whether radial migration is wholely responsible for sculpting the thick disk, long-lived perturbations initiated by flyby interactions could enhance radial migration after the two galaxies have separated and are essentially evolving in isolation.


$m=2$ features like bars and two-armed spirals can also transport gas to the centers of galaxies where it may form stars and even fuel SMBH growth \citep{Mihos1994,Hernquist1995,Hopkins2010b}. Gas inflow and star formation has also been tied with the formation of young disky pseudo-bulges with Sersic indices $\lesssim2$ \citep{Kormendy2004,Laurikainen2007,Scannapieco2010} which may obey a different $M-\sigma$ relation than classical bulges \citep{Hu2008,Graham2008}. If flyby-induced $m=2$ features drive gas inflow, enhanced central star formation, AGN activity, or placement on the $M-\sigma$ relation could signify a recent flyby or evolution history dominated by flybys.

Flybys could also shut-off smooth gas accretion, quenching star formation in the secondary. Combined with tidal stripping, galaxies with massive companions should have higher mass-to-light ratios and redder colors. Since massive halos have more flybys~\citep{SH12}, this effect would be more prominent in groups and clusters. Such an excess in the red-fraction has been reported~\citep{WYMBK09} and \citet{WTCB13} recently speculated that excess in the red-fraction could be attributed to flyby galaxies outside the host virial radius.

\section{Summary \& Future Work}\label{S_summary}

Using N-body simulations, we investigated the ability of galaxy flyby interactions to form bars. We used observationally-based techniques to identify and measure bar properties. We find that:
\begin{itemize}
	\item \emph{Flybys can induce $m=2$ perturbations in galaxy disks} reaching 6\% of the total potential. 
	\item \emph{Flybys can trigger bar formation} with bar strength decreasing in the primary and increasing in the secondary as mass discrepancy increases.
	\item \emph{Strong flyby-induced bars can persist long after the interaction.} In the 1:1 and 10:1 prograde interaction, the bars formed persisted ($A_{m=2}>4$\%) until the end of the simulations at 5\unit{Gyr} ($\sim4$\unit{Gyr} after pericenter).
	\item \emph{Flybys can induce spiral arms.} Two-armed spirals were formed in both the primary and secondary galaxies during both prograde interactions and persisted to the end of the simulations.
	\item \emph{Planar retrograde flybys do not form bars or spiral arms.} 
\end{itemize}

We have shown that flybys can, in principle, transform galaxy morphology. However, the precise role of flybys in bar formation requires further research to explore the strength of flyby perturbations as a function of orbit eccentricity, inclination and impact parameter, as well as to characterize any bulk differences between this formation mechanism and one that is more secular. 
Since these initial simulations were purely collisionless, it is unclear how well flyby-induced bars can drive gas inflow to enhance star formation and/or fuel SMBH growth. Future work will focus on probing the interaction parameter space, including hydrodynamics, and combining results with statistics from cosmological simulations to predict bar formation rates.

We exclude bulges to maximize the effect of the flybys on each galaxy's disk. Because central mass concentrations like a stellar bulge are thought to stabilize disks against bar formation \citep{Shen2004,Athanassoula2005}, the results presented can be viewed as an upper limit on the ability of flybys to form bars in disk galaxies. For a given interaction, the strength of any $m=2$ mode can be expected to decrease with increasing bulge mass and concentration \citep{Athanassoula2005}. However, quantifying the exact effect that bulge properties have on $m=2$ modes induced by flyby interactions will require additional simulations.

\acknowledgments

K.H-B. acknowledges the support of a National Science Foundation Career Grant
AST-0847696, and a National Aeronautics and Space Administration Theory grant NNX08AG74G. 
Support for M.L. was provided by a National Science Foundation Graduate Research Fellowship.
Supercomputing support was provided by Vanderbilt's Advanced Center for Computation Research and Education (ACCRE) and the Stampede system at Texas Advanced Computing Center (TACC) through the Extreme Science and Engineering Discovery Environment (XSEDE), which is supported by National Science Foundation grant number OCI-1053575.



\begin{table*}[ht]
\begin{center}
\begin{tabular}{|l||c|c||c|c|}
\hline
Galaxy 					& \multicolumn{2}{|l||}{MW1} 					& \multicolumn{2}{|l|}{MW10}					\\\hline\hline
Component				& Disk 					& Halo					& Disk					& Halo 					\\\hline
Total Mass (\Msol)		& $4.0\times10^{10}$ 	& $1.2\times10^{12}$	& $4.0\times10^{9}$		& $1.2\times10^{11}$	\\\hline
Scale Radius (kpc)		& $4.9$					& $43$					& $2.3$					& $20$					\\\hline
Scale Height (kpc)		& $0.49$				& n/a					& $0.23$				& n/a					\\\hline
\# of Particles			& $5.0\times10^{5}$		& $9.4\times10^{6}$		& $5.0\times10^{4}$		& $9.4\times10^{5}$		\\\hline
Particle Mass $m_p$		& $8.0\times10^{4}$		& $1.3\times10^{5}$		& $8.0\times10^{4}$		& $1.3\times10^{5}$		\\\hline
Softening (kpc)			& $0.05$				& $0.058$				& $0.05$				& $0.058$				\\\hline
\end{tabular}
\end{center}
\caption{Summary of component properties for each model galaxy.}
\label{tab:components}
\end{table*}%


\end{document}